\documentclass[a4paper]{article}
\usepackage{amsmath,amsthm,amssymb, mathrsfs}

\usepackage{fullpage}
\usepackage{enumerate}
 
\usepackage{listings}
\usepackage{tikz-cd}
\usepackage{tikz}
\usepackage{comment}
\usepackage{mathtools}
\usepackage[english]{babel}
\usepackage[autostyle]{csquotes}
\usepackage[
backend=biber,
style=ieee]{biblatex}

\usepackage{hyperref}

\newcommand{\Co}{\mathbb{C}}
\newcommand{\Z}{\mathbb{Z}}
\newcommand{\N}{\mathbb{N}}

\newcommand{\M}{\mathscr{M}}

\def \r {\rightarrow}
\def \e {\epsilon}

\def \a {\alpha}
\def \lan {\langle}
\def \ran {\rangle}

\def \meet {\wedge}
\def \join {\vee}
\def \De {\Delta}

\newtheorem{theorem}{Theorem}[subsection]
\newtheorem{defin}[theorem]{Definition}
\newtheorem{lemma}[theorem]{Lemma}
\newtheorem{coro}[theorem]{Corollary}
\newtheorem{exmp}[theorem]{Example}
\newtheorem{remark}[theorem]{Remark}
\newtheorem{prop}[theorem]{Proposition}
\newtheorem{no_count_theorem}{Main Result}

\title{Generalizing Pauli Spin Matrices Using Cubic Lattices}
\author{Morrison Turnansky\\ \texttt{University of Virginia}
}
\begin{document}
\maketitle
\begin{abstract}
In quantum mechanics, the connection between the operator algebraic realization and the logical models of measurement of state observables has long been an open question. In the approach that is presented here, we introduce a new application of the cubic lattice. We claim that the cubic lattice may be faithfully realized as a subset of the self-adjoint space of a von Neumann algebra. Furthermore, we obtain a unitary representation of the symmetry group of the cubic lattice. In so doing, we re-derive the classic quantum gates and gain a description of how they govern a system of qubits of arbitrary cardinality. \newline 
\end{abstract}

\section*{Acknowledgements}
\thispagestyle{empty}
\markboth{Acknowledgements}{Acknowledgements}
I am grateful to Dr. J. S. Oliveira for introducing me to the topic of cubic lattices. Also, I would like to thank Professor B. Hayes for the helpful comments both for this paper and the thesis on which it is based. 
\tableofcontents

\section{Introduction}
The cubic lattice \cite{rota} has long been thought of as an analogue of the standard Boolean lattice when adapted to the indeterminate setting of quantum logic. With this in mind, we see a substantial amount of literature that has been produced outlining the properties of a potential logic whose states are the cubic lattice in the finite case  \cite{logic_n_cube}. On the other hand, \cite{oliveira} introduces an axiomatic description of the cubic lattice without cardinality restrictions. We aim to combine these results. In so doing, we will obtain observables of an infinite quantum system and re-derive a universal set of quantum gates in the sense of the Solovay–Kitaev theorem. The key insight is that the reflection symmetries under consideration here can be represented as a subgroup of the unitary operator, which will be utilized to create a novel operator based realization of a cubic lattice.

As none of the referenced approaches introduce an analytic structure, it is a natural starting point as we consider the infinite case. Therefore, we embed the cubic lattice into a specifically constructed Hilbert Lattice. 
\begin{no_count_theorem}[Theorem \ref{embedding}]
Let $H$ be a Hilbert space constructed as a tensor product of $2$ dimensional spaces over an index set $I$. For the given Hilbert lattice $HL$ of $H$, there exists a cubic lattice $CL$ such that $CL \subseteq HL$, and the atoms of $CL$ are projections onto subspaces $H$ forming an orthonormal basis of $H$.
\end{no_count_theorem}

\noindent As the Hilbert lattice is much larger than our cubic lattice, We consider the minimal von Neumann algebra containing $CL$ as well.   
\begin{no_count_theorem}[Theorem \ref{minimal cubic lattice}]
The atoms of $W^\ast(\{s_i\}_{i \in I})$ are the atoms of CL.
\end{no_count_theorem}

\noindent We proceed to describe the algebra in our embedding of the cubic lattice, and in so doing, we generalize the standard result that the Pauli matrices span $M_{2}(\Co)$.

\begin{no_count_theorem}[Theorem \ref{B(H) isomorphism cubic}]
$B(H) = W^\ast(\{Us_iU^\ast\}_{i \in I}, \{s_i\}_{i \in I})$.
\end{no_count_theorem}

\noindent As a consequence, we generalize the Pauli matrices to infinite systems of qubits in our choice of matrix units when considered as a representation of $M_2(B)$ as opposed to $M_2(\Co)$, where $B \cong I_2 \otimes B(H_{I-i})$ for an indexing set $I$.

\[U_{\De_i} =  \begin{bmatrix}
       0 & 1 \\
       1 & 0 \\
   \end{bmatrix}\text{, } 
   s_i =  \begin{bmatrix}
       1 & 0 \\
       0 & -1  \\
   \end{bmatrix}\text{, and }
   is_iU_{\De_i} =  \begin{bmatrix}
       0 & i \\
       -i & 0 \\
   \end{bmatrix}.\]

\subsection{Background and Definitions}

The standard approach for approach for describing the spin states of $n$ qubits is to consider a tensor product of the form $\otimes_{i=1}^n \Co^2$ creating of vector space dimension $2^n$. In this setting each pure state is represented by an orthonormal basis vector

\begin{defin}
Let $H$ be a Hilbert Space. We define the lattice of closed linear subspaces of $H$ to be the Hilbert Lattice henceforth referred to as $HL$. In this context, $u \join v = span\{u ,v\}$, and $u \meet v = span\{u\} \cap span\{v\}$. 
\end{defin}

In some literature the Hilbert lattice is referred to as a standard lattice. The term is used because this is the standard construction of lattice of projection operators of a Hilbert space, we refer the reader to \cite{geo} for an in depth discussion. We will call the lattice $HL$.

The major issue with the above approach is that the geometry of the state space is not preserved because the dimensionality is too large. There are many unitary transformations that violate physical meaning, so we need a more restrictive symmetry group. With this in mind, we now move to the cubic lattice.

\begin{defin}[\cite{oliveira}]\label{cubic lattice definition}
A cubic lattice, $C$, is a lattice with 0 and 1 satisfies the following axioms:
\begin{enumerate}
\item For $x \in L$, there is an order-preserving map $\Delta_x: (x) \r (x)$, $(x)$ denotes the principal ideal generated by $(x)$.
\item If $0 < a,b <  x,$ then $a \join \Delta_x(b) < x$ if and only if $a \meet b = 0$.
\item $L$ is complete.
\item $L$ is atomistic.
\item $L$ is coatomistic. 
\end{enumerate}
\end{defin}

In the finite case, the cubic lattice can be thought of as a lattice of the faces of an $n$-cube. For an arbitrary cardinal, the axiomatic description above relies upon antipodal symmetry. We now tie together the geometric notion of the faces of the $n$-cube to the lattice of signed sets. 

\begin{defin}
Let $S = \{1, 2, \dots, n\}$ a signed set on $S$ is a pair $x = (A^+, A^-)$ of subsets of $S$ such that $A^+ \cap A^- = \emptyset$. The collection of signed sets is denoted by $L^+(S)$ is a poset with order relation $\le$ defined by reverse inclusion $x = (A^+, A^-) \le y = (B^+, B^-)$ if and only if $B^+ \subseteq A^+$ and $B^-  \subseteq A^-$. The pair $(A^+, A^-)$ uniquely determines the face $F$ if $A^+ \cap A^- = \emptyset$.
\end{defin}

Now that we have considered a poset of the faces of a cube defined as a signed set, we can consider the lattice of signed sets. For some intuition from the finite case, the vertices of the cube are the atoms of the lattice and its respective signed set $(A^+, A^-)$ partitions the indexing set $I$. In contrast, the whole cube is represented by $(\emptyset,\emptyset)$. The ordering of the signed set can also be thought of as the inclusion respective sub-faces of the cube. 

\begin{defin}
If $F$, $G$ are faces of $I^n$ such that $F,G \ne \emptyset$, with $F = (A^+, A^-)$ and $G = (B^+, B^-)$, then $G \subseteq F$ if and only if $A^+ \subseteq B^+$ and $A^- \subseteq B^-$. 
Let $\mathscr{F}(I^n)$ be the set of all faces of $I^n$ ordered by the above notion, so that $\mathscr{F}(I^n)$ forms a complete lattice, where $\join$ is the union of faces, and $\meet$ is the intersection of faces. With the addition of a $0$ element, $L^+(S)$ becomes a lattice denoted by $L(S)$ where for $x,y \in L(S)$, $x \join y = (A^+ \cap B^+, A^- \cap B^-) \in L(S)$ and $x \meet y = (A^+ \cup B^+, A^- \cup B^-) \in L(S)$ if $B^+ \cap A^- = \emptyset = B^- \cap A^+$ or $x \meet y = 0 \in L(S)$ otherwise. 
\end{defin}

In addition to $\meet$ and $\join$, a cubic lattice has an additional operation:

\begin{defin}\label{cubic def}
Every cubic lattice $L(S)$, in addition to the operations $\join$, $\meet$ also admits a partially defined operation $\Delta: L(S) \times L(S) \r L(S)$ defined by $\Delta(x,0) = 0$, and if $0 < x = (A^+, A^-)$, $0 < y = (B^+, B^-)$, $y \le x$, then $\Delta(x,y) = (A^+ \cup (B^- - A^-), A^- \cup (B^+- A^+))$.
\end{defin}

We have now given a very terse description of cubic algebras, and we will now move towards creating a faithful realization of the cubic algebra as an operator algebra. A large amount of technology must be developed as we do not yet even have a linear space of operators with which to begin.


\section{Embeddings of the Cubic Lattice and Octehedral Lattice}
Now that we have introduced the basic structures, we can build the necessary embedding to demonstrate that cubic lattices have a realization as a von Neumann algebra. In addition, we discuss the algebraic structure of the Hilbert lattice and compare it to the poset structure of the cubic lattice. Lastly, we compare the dual spaces with respect to both spaces categories. We show that there is, in a reasonable sense, a direct relationship between the dual of the poset and the dual of the analytic structure. 

\subsection{Cubic Lattice as a subset of a Hilbert Lattice}
We adapt the following definitions and proposition from \cite{hlatt} to our notation.

\begin{prop}\label{hlat}
The Hilbert lattice is an atomic, (completely) atomistic, complete, orthomodular lattice. \cite{hlatt}
\end{prop}

For the following theorem, we will be constructing a Hilbert space from an infinite tensor product. We do so in an established but non-standard way. We outline the necessary definitions for expository purposes and use the results from \cite{VN2}. Unless otherwise stated, when we refer to a Hilbert space formed by infinite tensor products, we mean the following construction, not the standard construction. 

For the following, $I$ is an index set of not necessarily countable cardinality, $H_\alpha$ is a finite dimensional Hilbert space for all $\alpha \in I$, and the norm on $f_\alpha \in H_\alpha$ is the norm of the Hilbert space. 

\begin{defin}\label{convergent} \cite{VN2}
$\Pi_{\alpha \in I}z_\alpha$, $z_\alpha \in \Co$, $\alpha \in I$, is convergent, and $a$ is its respective value if there exists for every $\delta > 0$, a finite set $I_0 = I_0(\delta) \subseteq I$, such that for every finite set $J = \{\alpha_1, \dots, \alpha_n\}$ (mutually distinct $\alpha_i$) with $I_0 \subset J \subset I$
\[|z_{\alpha_1} \cdot \dots \cdot z_{\alpha_n} - a| \le \delta.\]
\end{defin}

\begin{defin}
$\Pi_{\alpha \in I} z_\alpha$ is quasi-convergent if and only if $\Pi_{\alpha \in I} |z_\alpha|$ is convergent. It value is 
\begin{enumerate}
    \item the value of $\Pi_{\alpha \in I} z_\alpha$ if it is convergent
    \item $0$, if it is not convergent.
\end{enumerate}
\end{defin}

Now that we have a looser notion of convergence for infinite products, we adapt these definitions to functions in a normed space.

\begin{defin}
A sequence $f_\alpha$, $\alpha \in I$, is a C-sequence if and only if $f_\alpha \in H_\alpha$ for all $\alpha \in I$, and $\Pi_{\alpha \in I} ||f_\alpha||$ converges. 
\end{defin}

As we have an inner product for each $H_\alpha$, we can consider the infinite product of the respective inner products.

\begin{lemma}
If $f_\alpha,$ $\alpha \in I$, and $g_\alpha$, $\alpha \in I$ are two C-sequences then $\Pi_{\alpha} \langle f_\alpha, g_\alpha \rangle$ is quasi-convergent. \cite{VN2}
\end{lemma}

\begin{defin}
Let $\Phi(f_\alpha; \alpha \in  I)$ be the set of functionals on the product $\Pi_{\a \in I} H_\a$ which is conjugate linear in each $f_\alpha \in I$ separately over C-sequences. The set of all such $\Phi$ for any C-sequence will be denoted by $\Pi \odot_{\alpha \in I} H_\alpha$. We note that $\Pi \odot_{\alpha \in I} H_\alpha$ is a linear space, but it is not an inner product space. 
\end{defin}

Although each functional, $f_\alpha$ is conjugate linear for its respective $H_\alpha$, we do not have an inner product on the entire space. We can form an conjugate linear inner product space by considering a fixed C-sequence.   

\begin{defin} \label{basis}
Given a C-sequence $f_\alpha^0$, $\alpha \in I$, we form the functional $\Phi(f_\alpha; \alpha) = \Pi_{\alpha \in I}(f_\alpha^0,f_\alpha)$ where $f_\alpha$, $\alpha \in I$ runs over all C-sequences. Denote such a functional by $\Pi \otimes_{\alpha \in I} f_{\alpha}^0$.
\end{defin}

We now turn the inner product space into a linear space.

\begin{defin}
Consider the set of all finite linear aggregates of the above elements:
\[\Phi = \sum_{v=1}^p \Pi\otimes_{\alpha \in I} f_{\alpha,v}^0\]
where $p = 0,1,\dots$, $p$ and $f_{\alpha,v}^0$, $\alpha \in I$ is a C-sequence for each $v = 1, 2, \dots, p$. Denote the set of these $\Phi$ by $\Pi'\otimes_{\alpha \in I} H_\alpha$. For $\Phi = \sum_{v = 1}^p \Pi \otimes_{\a \in I} f_{\a, v}^0$, $\Psi = \sum_{\mu = 1}^q \Pi \otimes_{\a \in I} g_{\a ,\mu}^0 \in \Pi'\otimes_{\alpha \in I} H_{\alpha}$ we define the inner product by: 
\[\langle \Phi, \Psi \rangle  = \sum_{v = 1}^p \sum_{\mu = 1}^q \Pi_{\alpha \in I} \langle f_{\alpha,v}^0, g_{\alpha,\mu}^0 \rangle.\]
\end{defin}

The Hilbert space of \cite{VN2} has an inner product defined by a specific decomposition. For completeness, we highlight that the inner product is well defined.

\begin{lemma}
Let $\Phi, \Psi \in \Pi'\otimes_{\alpha \in I} H_{\alpha}$. The value of $\lan \Phi, \Psi \ran$ is independent of the choice of their respective decompositions. \cite{VN2}
\end{lemma}

Lastly, \cite{VN2} creates a Hilbert space by defining the completion with respect to our notion of convergence. 

\begin{defin}
Consider the functions $\Phi \in \Pi \otimes_{\alpha \in I} H_\alpha$ for which a sequence $\Phi_1$, $\Phi_2$, $\dots \in \Pi'\otimes_{\alpha \in I} H_\alpha$ exists such that 
\begin{enumerate}
    \item $\Phi(f_\alpha; \alpha \in I) = \lim_{r \r \infty} \Phi_r (f_\alpha; \alpha \in I)$ for all C-sequences $f_\alpha,$ $\alpha \in I$,
    \item $\lim_{r,s \r \infty} ||\Phi_r - \Phi_s|| = 0$
\end{enumerate}
The set they form is the complete direct product of $H_\alpha$, $\alpha \in I$ to be denoted by $\Pi \otimes_{\alpha \in I} H_\alpha$. Note that $\Pi'\otimes_{\alpha \in I} H_\alpha \subseteq \Pi \otimes_{\alpha \in I} H_\alpha \subseteq \Pi \odot_{\alpha \in I} H_\alpha$.
\end{defin}

For our application, the convergence criteria of Definition \ref{convergent} is acceptable. We will only be concerned with forming the tensors of elementary basis elements of the respective $H_\alpha$, so all of our elements are functionals derived from C-sequences as in Definition \ref{basis}. We can then consider their span in the natural way.

Lastly, we want to highlight that the Hilbert space construction results is separable only if each $H_\alpha$ is finite dimensional and $|I|$ is finite. Therefore the Hilbert spaces we are considering will in general be non-separable. 

\begin{theorem}\label{embedding}
Let $H$ be a Hilbert space constructed as a tensor product of $2$ dimensional spaces over an index set $I$. For the given Hilbert lattice $HL$ of $H$, there exists a cubic lattice $CL$ such that $CL \subseteq HL$, and the atoms of $CL$ are projections onto subspaces $H$ forming an orthonormal basis of $H$. 
\end{theorem}

\begin{proof}
We begin with the standard construction of a basis over a tensor product of index $I$. Let $e_i^+$, $e_i^-$ represent the $2$ basis vectors for $i \in I$. 

We now have that each elementary tensor is C-sequence as each element $||e_i|| =1$, so we have a linear functional of the form in Definition \ref{basis} in $H$, and it can be represented by its respective projection operator. As these are projections onto 1 dimensional subspaces, they are atoms in $HL$, and in the cone $B(H)^+$.

For each atomic elementary tensor described above, we use the notation, $v = \{A^+, A^-\}$, where $A^+ = \{i \in I : v_i = e_i^+\}$ and $ A^- = \{i \in I : v_i = e_i^-\}$. By the construction of $v$, we have that $A^+ \cap A^- = \emptyset$, and $A^+ \cup A^- = I$. Now we observe that the all such $v$ form the atoms of a signed set over the indexing set I. 

We define $CL = L(S_I)$, the lattice of signed sets generated by the closure of the above atoms under the operations of meet and join from the definition of cubic lattices. Recall by Definition \ref{cubic def}, that $\Delta : L(S) \times L(S) \r L(S)$ can be defined on any signed set.

As we have a description of the atoms of the cubic lattice in $\M^+$, we need to show that the atoms are closed under $\join$. Consider $a,b \in CL \cap HL$, where $a = \{A^+, A^-\}$ and $b = \{B^+, B^-\}$. Then $a \join_{CL} b = \{A^+ \cap B^+, A^- \cap B^-\}$. We now have that  $a \join_{CL} b$ is the projection $P_V$ onto the subspace $V = \otimes_{i \in I} V_i$ where $V_i= e_i^+$ for $i \in A^+ \cap B^+$, $V_i = e_i^-$ for $i \in A^- \cap B^-$, and $V_i = span\{e_i^+, e_i^-\}$ otherwise, so that $a \join_{CL} b \in HL$ and $P_V \in \M^+$. Therefore $CL \subseteq HL$. In addition as any element of $CL$ is a join of its atoms by atomisticity, and $0 \in HL$ trivially the result follows. 

The atoms of $CL$ form an orthonormal system in $H$. For any distinct atoms $a$, $b \in CL$, we have that there exists $i \in I$ such that $a_i \ne b_i$, so $\langle a_i , b_i \rangle_{H_\alpha} = 0$ which implies that $\langle a ,b \rangle_H = 0$. Furthermore, these vectors span $\Pi'\otimes_{\alpha \in I} H_\alpha,$ and therefore are dense in $H$.
\end{proof}

\begin{remark}
The Hilbert lattice is not a cubic lattice. Suppose not, then there exists a signed set realization of $HL$, $L(S)$ \cite{oliveira}. Let $r(\cdot)$ denote the rank of a subspace.  Consider the join of two linearly independent atoms $a = \{A^+, A^-\}, b = \{B^+, B^-\}$ such that $|\{A^+ - B^+\}| > 1$, so $r(a \join_C b) = |\{(A^+ \cap B^+) \cup (A^- \cap B^-)\}| > 2$. Then  $2 = r(a \join_H b) < r((A^+ \cap B^+) \cup (A^- \cap B^-))  = r(a \join_C b)$.
\end{remark}



We now discuss relations of the distinct lattice structures of the cubic lattice and Hilbert lattice. 

\begin{defin}
We say that $CL \subseteq B(H)$ and $H$ is constructed as in Theorem \ref{embedding} to mean the set of orthogonal projections onto their respective closed subspaces of $CL$ are in $B(H)$. We will use the notation: $a \in CL$, and $p_a \in B(H)$. 
\end{defin}

It is worth discussing why we chose to construct a non-separable Hilbert space. The standard approach to model a n-qubit system is to embed them into a $2^n$ dimensional space. In order to keep our later results consistent with this property, we are forced for an $|I|$-qubit system to embed into an $2^{|I|}$ dimensional space, which again is countable if and only if $|I|$ is finite. \\
We now explore how some of the operations of the cubic lattice and Hilbert lattice relate. 

\begin{coro}\label{Delta and perp agreement}
The action of $^\perp$ on $HL$ on the coatoms of $CL$ is a symmetry and coincides element-wise with the unitary symmetry associated with $\Delta$.
\end{coro}

\begin{proof}
The result follows as for all $c \in CL$, $p_c^\perp = 1 - p_c = p_{\Delta(c)} = U_\Delta p_c U_\Delta$. 
\end{proof}

\begin{defin}
Let $V = \otimes_{i \in I} V_i$ for some index set $I$ over vector spaces $\{V_i\}_{i \in I}.$ A generalized simple tensor of $V$ is a subspace of $V$ of the form $\otimes_{i \in I} U_i$, where $U_i$ is a subspace of $V_i$.
\end{defin}

\begin{coro}
The set of $CL \subseteq B(H)$ are exactly the operators represented by generalized simple tensors in the orthonormal basis. 
\end{coro}

Lastly, although the join operation differs on the cubic lattice and the Hilbert lattice, the meet operation is the same.

\begin{theorem}\label{meet agreement}
For a proper principal lattice filter of the cubic lattice, $F \subseteq CL\subseteq HL$, $\meet_H : F \times F  \r HL = \meet_C : F \times F \r F$.  Equivalently the join of and ideal OL agrees with meet of the Hilbert lattice.
\end{theorem}

\begin{proof}
Let $a, b \in F$. By definition we can write $a$, $b$ as the joins of atoms that are members of the orthonormal basis constructed in Theorem \ref{embedding}, so we can write $a \meet_H b$ in the same orthonormal basis as well. Therefore, we have the same relevant set of atoms for both $HL$ and $CL$ and reduce to this case implicitly for the remainder of the proof.

If $\alpha$ is an atom of $CL\subseteq HL$ such that $\alpha \le a$ and $\alpha \le  b$ then $\alpha \le a \meet_C b$ and $\alpha \le  a \meet_H b$. In addition, these are the only atoms in the commutative Boolean sub-lattice of $HL$ that are less than or equal to $a \meet_C b$ or $a \meet_H b$. By atomisticity of the cubic lattice and the Boolean sub-lattice of the Hilbert lattice,  $a \meet_C b = \join_C\{\a : \alpha \le a \text{ and } \alpha \le b\}$, $a \meet_H b = \join_H\{\a : \alpha \le a \text{ and } \alpha \le b\}$.

As the ordering of $CL$ is inherited from $HL$, $\alpha \join_{H} \beta = inf\{c \in HL: c \ge \alpha, \, c \ge \beta\} \le inf\{c \in CL: c \ge \alpha, \, c \ge \beta\}$. Therefore, $a \meet_H b \le a \meet_C b$. Now by reversing the above argument,  $\alpha \meet_{H} \beta = sup\{c \in HL: c \le \alpha, \, c \le \beta\} \ge sup\{c \in CL: c \ge \alpha, \, c \ge \beta\}$, and $a \meet_H b \ge a \meet_C b$. 
\end{proof}

\subsection{The Lattice Dual as an Algebra Anti Isomorphism.}

\noindent In order to expand our discussion of $CL$ and $HL$ as sets, we would benefit from compactness. Therefore, we consider the pre dual space $\M_*$ and dual $\M^*$ of $\M.$

\begin{defin}[Definition 3.24 \cite{alfsen}]
Let $\sigma, \omega \in \M_*^+$, where $\M$ is a von Nuemann algebra. We say that $\sigma$ is absolutely continuous with respect to $\omega$, written as $\sigma << \omega$, if $\sigma(q) = 0$ for all projections $q \in M$ such that $\omega(q) = 0$. 
\end{defin}

\begin{theorem}[Theorem 3.27 \cite{alfsen}]
If $\M$ is a von Neumann algebra and $\omega \in M_\ast^+$, then the norm closure of the face generated by $\omega \in M_\ast^+$ consists of all $\sigma \in M_\ast^+$ such that $\sigma << \omega$. 
\end{theorem}

\begin{prop}\label{face isomorphism}
For a base norm space X with generating hyperplane K, there is an order isomorphism from the non-zero faces of X to the faces of K.
\end{prop}

This is a standard fact, where the morphism is defined by a face $F$ in $X$ induces a face $F \cap K$ in $K$. One can also see this as a map from $0 \ne x \in X$ to $x / ||x||$ assuming X is a normed space and observing the induced facial structure.

\begin{prop} \label{normal state space faces}
The self adjoint part $M_*^+$ of the predual of a von Neumann algebra $\M$ is a base norm space whose distinguished base is the normal state space $K_*$ of $\M$. \cite{alfsen}
\end{prop}  

\begin{prop}\label{predual face to normal state space face}
If $F$ is a face in $\M_*^+$, then there is an order isomorphism to faces in the normal state space $K_*$.
\end{prop}

\begin{proof}
A direct result of Proposition \ref{normal state space faces} and Proposition \ref{face isomorphism}. 
\end{proof}

We use a direct application of \cite{alfsen} with slight abbreviation to avoid introducing notation that we will not use. For the full statement see references.  

\begin{prop}[\cite{alfsen} Theorem 3.35]\label{Alfsen Theorem 3.35}
\label{antiismorphism}
Let $\M$ be a von Neumann algebra with normal state space $K_*$, and denote $\mathscr{F}$ the set of all norm closed faces of $K_*$, by $\mathscr{P}$ the set of all projections in $\M$, and by $\mathscr{J}$ the set of all $\sigma$-weakly closed left ideals in $\M$, each equipped with the natural ordering. Then there is an order preserving bijection $\Phi: p \r F$ from $\mathscr{P}$ to $\mathscr{F}$, and an order reversing bijection $\Psi: p \r J$ from $\mathscr{P}$ to $\mathscr{J}$, and hence also an order reversing bijection $\Theta = \Psi \circ \Phi^{-1}$ from $\mathscr{F}$ to $\mathscr{J}$. The maps $\Phi$, $\Psi$, and $\Theta$ and the final inverse are explicitly given by the equations

\begin{enumerate}[(i)]
\item $F = \{\sigma \in K_* | \sigma(p) = 1\}$, 
\item $J = \{a \in \M | ap = 0 \}$
\item $J = \{a \in \M | \sigma(a^*a) = 0$ all $\sigma \in F\}$, $F = \{\sigma \in K_* | \sigma(a^* a) = 0$ all $a \in J\}$
\end{enumerate}
\end{prop}

We now want to show how our geometrically inspired $\De$ can be used somewhat synonymously with $\perp$ even across the dual space.  We first have to embed the lattice pre-dual, the octehedron, into the predual of our von Neumann algebra $B(H)$, where $H$ is constructed as in Theorem \ref{embedding}. Note that the lattice dual is reflexive, so the dual and predual are equivalent in this context. We first introduce some simplifying notation.
\begin{defin}
In the higher dimensional embedding, we lose the $+1$, $-1$ directionality to gain orthogonality. Therefore each $i \in S+$ and $j \in S^-$  corresponds to a mutually linearly independent linear functional for a total of $2|S|$ linear functionals. As an example let $j \in A^+$, and $f_i \in \{e_i^+, e_i^-\}$, and $p_{\otimes_{i \in S} f_i}$ for all $i \in S$ be the projection onto $\otimes_{i \in S} f_i$, then 
\[\e_j(p_{\otimes_{i \in S} f_i})
= \begin{cases}
1 & f_j = e_j^+\\
0 & f_j = e_j^-
\end{cases}\]
and extend linearly. 
\end{defin}

\begin{defin}
Define $\phi: CL \r \mathscr{F}$ by $\phi((A^+, A^-))$ as the norm closed convex hull of the linear functionals $\{\e_i: i \in A^+\} \cup \{\e_j: j \in A^-\}$. 
\end{defin}

As we will show the above $\phi$ will be the the analytic equivalent of our $\phi$ defined as a lattice anti isomorphism, and it will agree on the corresponding lattices, so the reuse of notation is intentional. 

\begin{defin}
We define a unitary operator denoted $U_{\Delta}$ by linearly extending its action on the basis of $H$, and letting $U_\De$ act by inner automorphism on orthogonal projections of subspaces of $HL$.
\end{defin}

As will be relevant later $\De(a,b)$ is linearly extendable in this representation of $CL$ if and only if $a = 1$. We now embed the octehedron into the pre-dual. 

\begin{lemma} \label{theta inverse is phi and delta}
Let $CL \subseteq HL$ as in Theorem \ref{embedding} with corresponding projections in $B(H)$. Then the restriction of the anti-isomorphism $\Theta^{-1}: \mathscr{J} \r \mathscr{F}$ of Proposition \ref{antiismorphism} to CL is equal to $\phi \circ U_\Delta: CL \r \mathscr{F}$. 
\end{lemma}


\begin{proof}
Let $J$ be the left ideal generated by a projection operator, $p_{(A^+, A^-)}$, onto a subspace of $(A^+, A^-) \in CL \subseteq HL$.  In addition, $U_\De p_{(A^+, A^-)} U_\De = p_{\De(A^+, A^-)}$. For simplicity, we will assign $(B^+, B^-) = \De(A^+, A^-)  = (A^- , A^+)$.

We claim that the face $\phi( U_\Delta p U_\De)$ in the normal state space, $F_{U_\Delta(p)}  = \overline{\{\sigma \in K_\ast: \sigma << \phi( U_\Delta p U_\De)\}}^{||\cdot||}$ is equal to $\Theta^{-1}(J)$. Firstly we have for any state $\omega \in \phi( U_\Delta p U_\De)$, $\omega(p) = 0$ as $supp(\omega)$ is orthogonal to $p$. Therefore, $F_{U_{\De}(p)} \subseteq \Theta^{-1}(p)$. 

Suppose $F_{U_\De(p)} \subset \Theta^{-1}(p)$, and there exists a state $\gamma \in \Theta^{-1}(J)$ such that $\gamma$ is not absolutely continuous with respect to $F_{U_{\De(p)}}$. In particular, $\gamma$ is not absolutely continuous with respect to a subset of $F_{U_{\De(p)}}$, namely the extreme points of $F_{U_{\De(p)}}$, consisting of $\{\e_{i}: i \in B^+ \}\cup \{\e_j: j \in B^-\}$, so there exists some projection $a \in \M$ such that $0 \ne a \subseteq (\cap_{i \in B^+} Ker(\e_i)) \cap ( \cap_{j \in B^-} Ker(\e_j))$ and $\gamma(a) \ne 0$. By construction, any projection in $(\cap_{i \in B^+} Ker(\e_i)) \cap ( \cap_{j \in B^-} Ker(\e_j))$ is less than or equal to p, so $\gamma(p) \ne 0$, which is a contradiction.
\end{proof}

Just as the atoms of the cubic lattice corresponded to atoms of the Hilbert lattice, the co-atoms of the octehedral lattice, which are the image of the atoms of CL under the dual map, correspond to co-atoms of the lattice of faces of the predual.

\begin{theorem}
For the normal state space $K_*$ of $B(H)$ where $H$ is constructed as in Theorem \ref{embedding}, there exists an $OL$ such that the coatoms of $OL$ are contained in the coatoms of $K_*$.
\end{theorem}

\begin{proof}
By Theorem \ref{embedding}, the atoms of the cubic lattice form an orthonormal basis of $H$, and the map $\phi: CL \r OL$ as defined Lemma \ref{theta inverse is phi and delta} is an order reversing map. As $\phi$ is the restriction of the map in $M_*^+$ whose facial structure is equivalent to $K_*$, we have our result.
\end{proof}


\begin{exmp}
The above results do not hold for the coatoms of CL. For a 2 cube, we see that coatoms are rank 2 projection operators onto a given half space while the coatoms of the respective Hilbert lattice must be rank 3 operators. 
\end{exmp}

\section{The Necessity of the Cubic Lattice}
Throughout this document, we have created a sufficient structure to characterize the algebraic relations of an $|I|$-qubit system when considered in an analytic space. However, we now raise the question, what other structures suffice? Is there perhaps an entire set of such objects and what is the underlying characterizing feature? We now demonstrate that the symmetries required for an $|I|$-qubit system require a cubic lattice structure. Furthermore, we show that these algebraic relations are in fact measurable in the sense of \cite{weaver}. We also show which von Neumann algebras contain a cubic lattice of a given cardinality up to $\ast$-isomorphism.  

We will think of the commutant of $U_\De$ is in some sense generated by the automorphism group of the lattice of signed sets. We will discuss this more in the following section. 

\subsection{The Symmetry Group of the Cubic Lattice and Quantum Relations}
In the finite case, the automorphism (symmetry) group of the cubic lattice is the Coxeter group $B_n$ otherwise known as the hyperoctahedral group $O_n$.

\begin{defin}
Let $Per(C)$ be the group of permutations of coatoms of $CL$, $Per_\De(C)$ be the centralizer of $\De$ in $Per(C)$, and $L(S)$ the lattice of signed sets over $S$.
\end{defin}

\begin{theorem}\label{automorphism group of L(S))}
For a cubic lattice of cardinality $\aleph$, $L(S)$, $Aut(L(S)) \cong Per_\De(C) \cong  \Z_2 \wr S_\aleph $, where $\wr$ denotes the unrestricted wreath product. \cite{turnansky}
\end{theorem}


In \cite{turnansky}, their choice of embedding space is a Banach space of dimension equal to the indexing set S, as opposed to our exponentially larger Hilbert embedding. We now generalize these arguments to von Neumann algebras over the Hilbert space constructed in Theorem \ref{embedding}. 

\begin{prop}
The $C^*$ algebra generated by $U_\Delta$ is a von Neumann algebra. 
\end{prop}

\begin{proof}
Since $U_\Delta$ is a self adjoint unitary operator, we have that $C^\ast(U_\De)$ is a finite dimensional algebra, and so equal to its WOT closure. 
\end{proof}

\begin{lemma}
Let ' denote the commutant. Then $W^*(U_\Delta) = Z(W^*(U_\Delta)')$. 
\end{lemma}

\begin{proof}
As $W^*(U_\Delta)$ is an abelian unital $W^*$ algebra, $W^*(U_\Delta) \subseteq Z(W^*(U_\Delta)')$. Since $W^*(U_\Delta)$ is also a von Neumann algebra, we have that $Z(W^*(U_\Delta)') \subseteq W^*(U_\Delta)'' = W^*(U_\Delta)$, where the last equality follows by the double commutant theorem.
\end{proof}

We now have a large amount of insight into the structure $W^*(U_\Delta)$. There are three views to consider, firstly as a finite dimensional abelian von Neumann algebra $W^*(U_\Delta)$ is isomorphic to an $l^\infty(\{1,2,\dots, n\})$ for $n \in \N$. On the other hand, we know that $W^*(U_\Delta)$ as a unital commutative Banach algebra, so $W^*(U_\Delta)$ is also isomorphic to $C(K)$, and lastly that $W^*(U_\Delta)$ is isomorphic to $p(U_\Delta)$, which as $\Delta$ is an involution, is a two dimensional $\Co$ vector space. Of course, this all ultimately follows from the general principle that continuous maps over a finite space are a vacuous concept and devolves to a map from a finite set to the complex numbers. We describe this in general in the following statement.

\begin{prop}\label{finite order commutant}
Let $A \in B(H)$ be a normal operator such that $A^n = I$ for some $n \in \N$, then $C^*(A)$ is a von Neumann algebra and equal to the center of its commutant. 
\end{prop}

Before further discussing $W^\ast(U_\De)'$, we introduce some theory about the automorphism groups of the Hilbert lattice and the cubic lattice. To begin, the embedding of $L(S)$ in $H(L)$ as constructed in Theorem \ref{embedding} is minimal in a fairly strict sense. 

\begin{theorem} \label{universal}
Let $f: L(S) \r HL$ where the atoms of $L(S)$ are contained in the atoms of $HL$, and $f$ is an injective order morphism. Then there exists an unique injective order morphism $\psi:\tilde{H}(L) \r HL$, where $\tilde{H}(L)$ is the embedding, $j$ of Theorem \ref{embedding} such that $\psi \circ j = f$
\end{theorem}

\begin{proof}
First we show existence of such an $f$. We only need to use that the Hilbert lattice is atomic and complete by  Proposition \ref{hlat}. Therefore, if we have two Hilbert lattices, we have an injective order morphism if we have an injective mapping of orthonormal bases to the Hilbert spaces of the respective Hilbert lattices. Then we have $\psi = f \circ j^{-1}$. Now we see that uniqueness follows as $j$ is a bijection between the orthonormal basis of $\tilde{H}(L)$ and the atoms of $L(S)$, so if there exists another map $\rho$ satisfying our criteria, then $\rho \circ j = f$ implies $\rho  = f \circ j^{-1} = \psi$. 
\end{proof}

In order to study a representation of the automorphisms of the cubic lattice, we first look at representing automorphisms of the Hilbert lattice.

\begin{defin}
A conjugate linear operator is a linear operator except for scalar multiplication is treated as conjugate scalar multiplication. 
\end{defin}

\begin{defin}
Let $H$ be a Hilbert space and consider $\Phi: B(H) \r B(H)$. $\Phi$ is said to be implemented by a (conjugate) unitary if there is a (conjugate) unitary map $U: H \r H$ such that $\Phi a = UaU^\ast$ for all $a \in B(H)$. \cite{alfsen}
\end{defin}

\begin{lemma}\label{order automorphism are unitary}
Let $g \in Aut(H(L))$, then there exists a unitary or conjugate linear unitary operator $U_g: H \r H$ such that $g$ is implemented by $U_g$.
\end{lemma}

\begin{proof}
If $g \in Aut(H(L))$, then $g$ is a unital order automorphism, so by \cite[Proposition 4.19]{alfsen}, $g$ is a Jordan automorphism. If $g$ is a Jordan automorphism, then $g$ is either a $\ast$-isomorphism or $\ast$-anti-isomorphism by \cite[Proposition 5.69]{alfsen}. If $g$ is a $\ast$-isomorphism then $g$ is implemented by a unitary, and if $g$ is a $\ast$-anti isomorphism then $g$ is implemented by a conjugate unitary by \cite[Theorem 4.27]{alfsen}.
\end{proof}


\begin{defin}
Let $A \in \M$ be invertible, then $Ad_A: \M \r \M$ is defined by $Ad_A(\cdot) = A(\cdot)A^{-1}$. Equivalently, one can view $Ad_A$ as the inner automorphism induced by $A$ on $\M$. \cite{alfsen} 
\end{defin}

\begin{defin}
We say that the action of two unitary operators commute in a von Neumann algebra $\M$ if their action by inner automorphism commutes. 
\end{defin}

\begin{theorem}\label{actions commute}
Let $g \in Aut(H(L))$, then $Ad_g \in Aut(L(S))$ if and only if the action of $g$ commutes with the action of $U_\De$ on $W^\ast(L(S))$ where $H$ is constructed in the manner of Theorem \ref{embedding}. 
\end{theorem}

\begin{proof}
By Lemma \ref{order automorphism are unitary}, we know that $g$ can be implemented by a unitary or conjugate unitary operator, $U$. Without loss of generality, we assume that $U$ is a unitary operator as this affects the associative multiplication consistent with the Jordan algebra of the Hilbert lattice, but it does not affect the action as an order automorphism. 

Assume that the action $U$ commutes with action of $U_\De$ on $L(S)$. It is sufficient to show that $Ad_g \in Per_{\De}(C)$. Now $c$, $\Delta(c)$ be coatoms in $L(S)$. Then 
\[\Delta(g(p_c)) = U_\Delta U_g p_c U_g^\ast U_\Delta = U_g U_\Delta p_c U_\Delta U_g^\ast. = g(\Delta(p_c))\]

We have that $g$ maps to coatoms to coatoms in some isomorphic lattice to our original $L(S)$ in particular an order isomorphism, so $Ad_g \in Aut(L(S))$.

Now for the converse. If the inner automorphisms do not commute on $L(S)$, then there exists $c \in C$ such that $\Delta(g(p_c)) \ne g(\Delta(p_c))$. As $g \in Aut(L(S))$, $g(c)^\perp = \De(g(c))$ by Theorem \ref{Delta and perp agreement}. Therefore $g(\De(p_c)) \ne g(p_c)^\perp$, but by linearity $g(\De(p_c)) + g(p_c) = g(\De(p_c)  + p_c) = g(I) = I$, which leads to a contradiction.
\end{proof}

One can observe that there are many more unitary transformations, and therefore, automorphims of the Hilbert lattice, then there are automorphisms of the cubic lattice. Now that we know that $Aut(H(L))$ are unitary or conjugate unitary transformation, we deduce exactly which unitary operators are automorphisms of the cubic lattice.




We see that we have a choice of equivalence class when we represent $Aut(L(S))$ by its action on $L(S)$, as there are  automorphisms acting as the identity on $L(S)$ that to do not act as the identity on $HL$. Namely, the abelian von Neumann algebra, $W^\ast(\{p_c\}_{c \in C})$ i.e. the symmetries associated $L(S)$, are such an example. 

Due to this ambiguity, we choose to define a group representation of Aut(L(S)) up to group isomorphism acting on $H$ as opposed to inner automorphisms acting on $HL$. From another perspective, we have for each $U \in Aut(L(S))$, the action $Uh \ran \lan Uh_1 \mapsto h_2\ran \lan h_2$ on the lattice of orthogonal projections, $HL$, and we are instead considering the action $Uh_1\ran \mapsto h_2 \ran$ where $h_1, h_2 \in H$. One can see that any action in $Aut(L(S))$ can be induced by the action on $H$ and vice versa, but we have removed the ambiguity of the representation. We make this more formal below.   

\begin{lemma}\label{automorphism of L(S) are in the commutant}
There exists a unitary representation $\rho: Aut(L(S)) \r B(H)$  such that $U_g \in W^\ast(U_\De)'$ for all $g \in Aut(L(S)).$
\end{lemma}

\begin{proof}
When considered as an automorphism group acting on the orthonormal basis constructed in Theorem \ref{embedding}, we have a group representation of $Aut(L(S))$ contained in the permutation group over $H$, so we conclude that the group representation is a unitary representation. 

Now we apply $Aut(L(S)) \cong Per_{\Delta}(C)$,
the permutations of the coatoms that commute with $\Delta$ to see that $\De \subseteq Z(Aut(L(S))$. As commutativity of the group implies commutativity of its representation, we conclude the result. 
\end{proof}


We want to decompose $W^(U_\De)'$ into automorphisms of the cubic lattice and projections onto the cubic lattice. We first prove some facts about the maximality of an abelian algebra characterized by its projections. 

\begin{prop}\label{stonean completion}
Every complete Boolean algebra, $\mathscr{B}$, corresponds to a unique Stonean completion $\mathscr{A}$ whose set of projections is equal to $\mathscr{B}$. \cite{stonean}
\end{prop}

\begin{prop}\label{maximal abelian}
If $\mathscr{A} \le B(H)$ is an atomic abelian von Neumann algebra whose lattice of projections form an atomic complete Boolean algebra, which is maximal in the Hilbert lattice of $H$, then $\mathscr{A}$ is a maximal abelian algebra. 
\end{prop}

\begin{proof}
From Proposition \ref{stonean completion}, we know that Boolean lattice of projections correspond to abelian sub-algebras of a von Neumann algebra. Let $A$ be the atoms of $\mathscr{A}$, and $p \in  \mathscr{A}' - \mathscr{A}$ be a projection. Furthermore, we can assume that $p$ is orthogonal to every $a \in A$, otherwise let $p = p - (\join_{a \in A} p \meet a)$. Then $p$ commutes the atoms $a \in \mathscr{A}$, so $a \ge a \meet p \ge = ap = 0$ as $a$ is an atom. The Hilbert lattice is atomic by Proposition \ref{hlat}, so let $b \le p$ be an atom and by the above $a \meet b \le a \meet p = 0$.Therefore, the lattice containing $\mathscr{B}$  and $b$ is an atomic complete Boolean lattice strictly containing $\mathscr{B}$, contradicting maximality. Therefore $\mathscr{A}$ and $\mathscr{A}'$ contain the same projections and must be equal. The result follows as abelian von Neumann algebra equal to its commutant is maximal. 
\end{proof}

\begin{lemma}\label{choice of representation}
Let $CL \subseteq HL$ as in Theorem \ref{embedding} and $U\in W^\ast(\De)'$ be unitary. There exists a unitary $V \in \rho(Aut(L(S)))$ such that $Ad_U = Ad_V : CL \r CL$ and $U = VS$ for $S \in W^\ast(\{p_c\}_{c \in C}) \cap W^\ast(U_\De)'$. 
\end{lemma}

\begin{proof}
If $U \in W^\ast(\De)'$, then $Ad_U \in Aut(L(S))$ by Theorem \ref{actions commute}. Now let $V = \rho(Ad_U) \subseteq W^\ast(U_\De)'$. Then $Ad_{V^\ast} = Ad_V^{-1}$, so $Ad_{UV^\ast}|_{CL} = Ad_I|_{CL}$. As the action of inner automorphism stabilizes $CL$, $UV^\ast \in W^\ast(\{p_c\}_{c \in C})'$ and $W^\ast(\{p_c\}_{c \in C})'= W^\ast(\{p_c\}_{c \in C})$ by Proposition \ref{maximal abelian}.

Therefore, there exists $S \in W^\ast(\{p_c\}_{c \in C})$ such that $U = VS$. Furthermore, $S = UV^\ast$, so $S \in W^\ast(U_\De)'$ as well. 
\end{proof}

The above representation when considered as an action of inner automorphism on $B(H)$ can be seen to be identical to our previous notion where we fix $S = I$.


\begin{theorem}
$W^\ast(U_\De)' = W^\ast(\rho(Aut(L(S))), W^\ast(\{p_c\} , U_\De)')$. 
\end{theorem}

\begin{proof}
We use the above lemmas, to show both von Neumann algebras have the same set of unitaries for an appropriate representation of $Aut(L(S))$. In particular any unitary in $W^\ast(U_\De)'$ is a product of the two algebras $W^\ast(\rho(Aut(L(S))))$ and $W^\ast(\{p_c\} , U_\De)'$. Now we use that von Neumann algebras are generated by their unitaries, see Proposition I.4.9 in \cite{takesaki1}, so the result follows.  
\end{proof}

\begin{coro}
$W^\ast(U_\De) = Z(W^\ast(\rho(Aut(L(S)))))$.
\end{coro}

\begin{proof}
Follows immediately from the result $Z(Aut(L(S)) = \{1, \De\}$ in \cite{turnansky}, the definition of $\rho$, and the spectral theorem.
\end{proof}

We have  extended the purely group theoretic ideas of \cite{turnansky} to the more general von Neumann algebra setting. Now that we have a legitimate and well understood von Neumann algebra, $W^\ast(\De)$, and some insight into its commutator, we can finally discuss the quantum relations that $\Delta$ induces. We demonstrate that the relations specified by the cubic lattice are natural and measurable in the sense of \cite{weaver}. We first define a standard relation:

\begin{defin}
Let $X$ be a set, then a binary relation is a set of ordered pairs $(a,b) \in X \times X$, where $a,b\in X$. In some literature, the notation $aRb$ is used to describe a set with a relation $R$, denoted $(X,R)$.
\end{defin}

The obvious issue with the classic notion of a relation is that when one considers a non-atomic measure, these finite relations become vacuous. In \cite{weaver}, they generalize this notion to a measurable relation. 

\begin{defin}
A measure space $(X, \mu$) is finitely decomposable if $X$ can be partitioned into a (possibly uncountable) family of finite measure subspaces $X_\lambda$ such that a set $S \subseteq X$ is measurable if and only if its intersection with each $X_\lambda$ is measurable, in which case $\mu(S) = \sum_\lambda \mu(S \cap X_\lambda)$.
\end{defin}

As pointed out in \cite{weaver}, a measure space $(X, \mu)$ is finitely decomposable exactly when $L^\infty(X, \mu)$ is an abelian von Neumann algebra. A full explanation can be seen in \cite{abelianvna}.

\begin{defin}
Let $(X, \mu)$ be a finitely decomposable measure space. A measurable relation on $X$ is a family R of ordered pairs of nonzero projections in $L^\infty(X, \mu)$
such that $(\join p_\lambda, \join q_\kappa) \in R$ if and only if some $(p_\lambda ,q_\kappa) \in R$ for any pair of families of nonzero projections $\{p_\lambda\}$ and $\{q_\kappa\}$. \\ 
Equivalently, we can impose the two conditions\\
$p_1 \le p_2$, $q_1\le q_2$, $(p_1, q_1) \in R \Rightarrow (p_2, q_2) \in R$\\
and \\
$(\join p_\lambda, \join q_\kappa) \in R \Rightarrow$  some $(p_\lambda, q_\kappa) \in R$. \cite{weaver}  
\end{defin}

Of course we are dealing with a more general not necessarily abelian structure. In \cite{weaver}, they define a quantum relation on a von Neumann algebra.

\begin{defin}
A quantum relation on a von Neumann algebra $\M \subseteq B(H)$ is a $W^\ast$-bimodule over its commutant $\M'$,i.e., it is weak$^*$ closed subspace of $V \subseteq B(H)$ satisfying $\M'V\M' \subseteq V.$
\end{defin}

Now we argue from the reverse perspective. If we a priori argued that a quantum logic must respect the symmetry group of a possibly infinite dimensional cube, the infinite hyperoctahedral group, $Aut(L(S))$, we could consider the von Neumann algebra generated by $Aut(L(S))$. 

\begin{prop}\label{transitive action}
Let $\mathscr{B}$ be the basis of atoms on $L(S)$ constructed in Theorem \ref{embedding}, then $W^\ast(U_\De)$ acts transitively in $\mathscr{B}$.
\end{prop}

\begin{proof}
If $u,v \in \mathscr{B}$, then we consider the composition of $\Z_2$ actions on each disagreeing index, which is contained in $W^\ast(U_\De)$ by construction. 
\end{proof}

Geometrically in the finite dimensional case, we can observe that the Coxeter group, $B_n$, is the group of rigid motions of the cube and must be able to permute any 2 vertices. 



\begin{coro}
The quantum relations associated with $W^\ast(U_\De)$ are weak$^\ast$ closed subspaces $V$ satisfying $W^\ast(U_\De) V W^\ast(U_\De) \subseteq V$. 
\end{coro}

We now have that the operator systems discussed above, i.e. the ideals of $CL$ are have well defined quantum relations. Furthermore, by presupposing the lattice, we have re-derived both $\Delta$, and the invariant subspaces of the cube. 

From an experimental setting, this invariant subspace is a natural requirement, as one can rotate the axis for detection of a spin $\frac{1}{2}$ particle, but we still need cubic symmetry as the experiment takes place in Euclidean space. Therefore, our notion $\Delta$ can be viewed as a necessary condition of relations in the experiment. In addition, these symmetries can be verified by the single relation $\De$ as opposed to the (infinite) hyperoctahedral group. 

Furthermore, our original lattice based definitions of the cube are now seen to be measurable in a much more general sense.

As the principle ideals of a cubic lattice are again cubic lattices, we can further infer that the principle ideals form von Neumann subalgebras and therefore operator systems.
\begin{defin}
An operator system is unital $\ast$-closed subspace contained in a unital $C^\ast$ algebra. 
\end{defin}

To be precise we present the following theorem. 
\begin{theorem}
Let $p_a \in CL \subseteq HL$ as in Theorem \ref{embedding}, and $\M = B(H)$. Then $p_a \M p_a$ forms a von Neumann sub algebra, and $[a]_C$ forms a Boolean lattice.
\end{theorem}

\begin{proof}
The fact that $p_a \M p_a \le \M$ is a standard result. We refer to \cite{oliveira} for the construction of the complement $^c(\cdot) = a \join_C \De \De(\cdot, a)$ making $[a]_C$ a Boolean algebra.
\end{proof}

Therefore, an element in a cubic lattice can be seen as a dividing line between a Boolean algebra and a von Neumann algebra. This echoes back to our notion that the projection operators of a cubic lattice detect the minimum entangled state containing the respective atoms and above that level of detection, elements have become disentangled and therefore Boolean.

\begin{coro}
Let $(p)$ be a principle ideal in CL. Then $C^\ast(\Delta_p)$ is a von Neumann algebra $pDp$. 
\end{coro}

\begin{proof}
This follows as the principle ideals of cubic lattice are themselves cubic lattices. 
\end{proof}

\subsection{Operator Algebras containing a cubic lattice}
We can see that the above results can can be generalized in a straightforward manner.

\begin{defin}\label{coatom defintion}
Let $C$ be the co-atoms of $CL$. Then for each $c \in C$, we get a symmetry in the canonical form of $p_c - p_{\De c}$. We denote the set by $\{s_i\}_{i \in I}$. 

Importantly, ith coordinate in the tensor product is equal to the matrix 
$s = \begin{bmatrix}
    1 & 0 \\
    0 & -1 
\end{bmatrix}$.
\end{defin}

\begin{lemma}\label{mutual commutant}
With our previous choice of representation, $\rho: Aut(L(S)) \r B(H)$, the mutual commutant of $U_\De$, $s_i$ is equal to $W^{\ast}( W^\ast(\rho(\Z_2 \wr S_{I-i})), W^\ast(\{p_c\}_{c\in C}) \cap W^\ast(U_\De)')$, again by $\wr$ we mean the unrestricted wreath product. 
\end{lemma}

\begin{proof}
By Lemma \ref{choice of representation}, we already have an explicit definition of the unitaries that commute with $U_\De$, so we only need to consider the subset that also commute $s_i$. 

We consider the elements of $Per_{\De}(C)$, that fix $p_{c_i}$, $p_{\De c_i}$. They are the permutations fixing the ith coordinate that commute with $U_\Delta$, so we have that it is again the infinite hyperoctahedral group but on one less coordinate or $\Z_2 \wr S_{J}$, where $|J| = |I| - 1$ or elements of $W^\ast(\{p_c\}_{c \in C}) \cap W^\ast(U_\De)'$.

Now the result follows by taking the WOT closure of the algebra generated by its unitary operators, which fully defines the von Neumann algebra again by Proposition I.4.9 \cite{takesaki1}. 
\end{proof}

\begin{theorem}\label{isomorphism of B(H)}
Let $H$ be constructed in the manner of Theorem \ref{embedding}, then $B(H) \cong M_2(B)$, where $B \cong I_2 \otimes B(H_{I-i})$.
\end{theorem}

\begin{proof}
Let $U_{\De_i}$ be the tensor product whose ith index is equal to $U_\De$'s ith index and $I_2$ elsewhere. We claim the following form matrix units for $B(H)$. 
\begin{align*}
    e_{11} &= \frac{I + s_i}{2}\\
    e_{12} &= \frac{(I + s_i)U_{\De_i}} {2}\\
    e_{21} &= \frac{U_{\De_i}(I+ s_i) }{2}\\
    e_{22} &= \frac{I - s_i}{2}
\end{align*}

We can directly compute that $e_{11} + e_{22} = I$, $e_{12} = e_{21}^\ast$, and $e_{ij}e_{kl} = \delta_{jk}e_{il}$. Therefore, $B(H) \cong M_2(B)$, where $B$ commutes with all of the matrix units, see Lemma 4.27 of \cite{alfsen3}.

Now we show $N = W^\ast(\{e_{ij}\}_{i,j \in \{1,2\}}) = W^\ast(U_{\De_i}, s_i)$.

Firstly, $U_\De \in N$,
\begin{align*}
U_{\De_i} &=\frac{U_{\De_i} + s_i U_{\De_i} - s_i U_{\De_i} + U_{\De_i}}{2}\\
&=\frac{U_{\De_i} + s_i U_{\De_i} + U_{\De_i} s_i + U_{\De_i}}{2} \\
&= \frac{(I + s_i)U_{\De_i}} {2} + \frac{U_{\De_i}(I+ s_i) }{2} \\
&= e_{12} + e_{21}
\end{align*}

Secondly, $s \in N$
\begin{align*}
s_i &= \frac{2s_i - I + I}{2}\\
&= \frac{I+ s_i}{2} - \frac{I - s_i}{2}\\
&= e_{11} - e_{22} 
\end{align*}

Therefore $W^\ast(U_\De , s_i) \subseteq N$. For the reverse containment,  the generators of $N$ are in the algebra generated by $U_\De$, $s_i$, so they are in the WOT closure of the algebra. 

Now we apply that $M_2(\Co)_i \otimes I_{I - i}  = W^\ast(U_{\De_i}, s_i)$, so that $N' = I_2 \otimes B(H_{I-i})$, where $H_{I - i} = \otimes_{j \in (I - i)} \Co^2$ in the manner of Theorem \ref{embedding}.
\end{proof}

\begin{exmp}
We see that in our choice of matrix units, we again obtain that 

\[U_{\De_i} =  \begin{bmatrix}
       0 & 1 \\
       1 & 0 \\
   \end{bmatrix}\text{, } 
   s_i =  \begin{bmatrix}
       1 & 0 \\
       0 & -1  \\
   \end{bmatrix}\text{, and }
   is_iU_{\De_i} =  \begin{bmatrix}
       0 & i \\
       -i & 0 \\
   \end{bmatrix}.\]

This is considered as a representation of $M_2(B)$ as opposed to $M_2(\Co)$. Of course if we reduce to the single qubit case, we have that the $B \cong \Co$ and only one choice of index for $s_i$, so our result is consistent. 
\end{exmp}
We relate the above construction to a more familiar general object. 
\begin{defin}\cite{alfsen}
A Cartesian triple is a set of operators $r$, $s$, $t$ in a von Neumann algebra such that 
\begin{enumerate}
    \item $r \circ s = s \circ t = t \circ r  = 0$.
    \item $Ad_r Ad_s Ad_t = I$. 
\end{enumerate}
\end{defin}

\begin{coro}
For any $s_i \in S$, the set $U_\De$, $s_i$, and $iU_\De s_i$ form a Cartesian triple in B(H). 
\end{coro}

\begin{proof}
Given our representation, the result follows from standard facts about Pauli matrices. 
\end{proof}

We can consider another von Neumann subalgebra of B(H). Namely, $W^\ast(\{s_i\}_{i \in I})$. 

\begin{lemma}\label{coatoms of a von neumann subalgebra}
Given our representation of $\De$, the coatoms of $CL$, $C$, are exactly a generating set of projections of $W^\ast(\{s_i\}_{i \in I})$.
\end{lemma}

\begin{proof}
We have that $C = \left\{\dfrac{1 \pm s_i}{2}\right\}_{i \in I}$ generates $\{s_i\}_{i \in I}$ and vice versa. Therefore, $W^\ast(C)$ generates the unitaries of $W^\ast(\{s_i\}_{i \in I})$, and therefore generates all of $W^\ast(\{s_i\}_{i \in I})$. 
\end{proof}

\begin{theorem}\label{minimal cubic lattice}
The atoms of $W^\ast(\{s_i\}_{i \in I})$ are the atoms of CL.
\end{theorem}

\begin{proof}
We have shown that the coatoms of CL are in $W^\ast(\{s_i\}_{i \in I})$, by Lemma \ref{coatoms of a von neumann subalgebra}. By coatomicity of CL, and Lemma \ref{meet agreement}, we have that the atoms of CL are contained in $W^\ast(\{s_i\}_{i \in I})$.

Now for the reverse direction, we consider the complete  lattice of projections $L$ generated by the canonical projections of $\{s_i\}_{i\in I}$. Here we mean complete in the sense of lattice theory not necessarily complete with respect to the norm and generated in the sense closure of meet and joins. As the canonical projections of $\{s_i\}_{i\in I}$ are exactly the coatoms of $CL$, we have that the atoms of $L$ are exactly the set of atoms of $CL$ by  Lemma \ref{meet agreement}, and in addition, $L$ is a complete lattice generated by an orthonormal basis and therefore Boolean.

In our specific application of Proposition \ref{stonean completion}, the atoms of $L$ form a maximal set of mutually orthogonal projections, and the subalgebra of bounded operators of $\mathscr{A}$,  $C^\ast(L)$, is abelian, so we have that $C^\ast(\{s_i\}_{i\in I}) = C^\ast(L)$ is a von Neumann algebra \cite[Remark 10.8]{stonean} whose atoms are the atoms of the cubic lattice. 
\end{proof}

Therefore, we now have a minimal von Neumann algebra containing the $CL = L(S)$ for a given $|S|$. Furthermore, we have shown that $W^\ast(\{s_i\}_{i \in I}) \le B(H)$, where $H$ is minimal as in Theorem \ref{universal}.

\begin{exmp}
When reducing the one qubit case, we see that $W^\ast(U_\De, s)$ contain the Pauli matrices, which are a $W^\ast$ algebra over $\Co$ generating all of $M_2(\Co)$, which is a well known result, as required. Furthermore, we have a unitary matrix $T \in M_2(\Co)$,
\[T = \frac{1}{\sqrt{2}}\begin{bmatrix}
1 & 1 \\
1 & -1\\
\end{bmatrix},
\]
which is a unitary similarity sending $s$ to $U_\De.$ We recognize this as the normalized Hadamard matrix.
\end{exmp}

We are now in a position to generalize the result that the Pauli spin matrices span $M_2(\Co)$.

\begin{defin}
We define $U = \otimes_{i\in I} T$.
\end{defin}





\begin{theorem}\label{B(H) isomorphism cubic}
$B(H) = W^\ast(\{Us_iU^\ast\}_{i \in I}, \{s_i\}_{i \in I})$.
\end{theorem}

\begin{proof}
We only need to show that $W^\ast(\{Us_iU^\ast\}_{i \in I}, \{s_i\}_{i \in I})' = W^\ast(\{Us_iU^\ast\}_{i \in I})' \cap W^\ast(\{s_i\}_{i \in I})' = Z(B(H)) = \Co I$. 

Suppose that $V$ is a unitary operator commuting with $U_\De$, then by Lemma \ref{choice of representation}, $Ad_V \in Aut(L(S))$, when considering its action by inner automorphism on $L(S)$. As $V$ commutes with each co-atom of $L(S)$, $V$ acts trivially on the co-atoms of $L(S)$, so by coatomisticity of L(S), $V$ acts trivially on $L(S)$. Then 
$V \in W^\ast(\{s_i\}_{i\in I})$. By symmetry, $V \in W^\ast(\{Us_iU^\ast\}_{i\in I})$. 

Consider canonical projections $p_i$ of $Us_iU^\ast$ and $q_i \in s_i$ for some fixed index $i \in I$. Then $p_i \meet q_i = \lim_{n \r \infty} (p_iq_ip_i)^n  = \lim_{n \r \infty} (\frac{1}{2}p_i)^n = 0$. By construction, any atom $a \in \{Us_iU^\ast\}$, a is bounded by a canonical projection of $Us_iU^\ast$, so we assume without loss of generality that $a \le p$, and by symmetry we assume $b \le q$. Then $a \meet b \le p \meet q = 0$. Therefore the atomistic Boolean lattice of projections associated with $\{Us_iU^\ast\}_{i \in I}$ and $\{s_i\}_{i \in I}$ have distinct sets of atoms.
By atomisticity, $W^\ast(\{s_i\}_{i\in I})$ and $W^\ast(\{Us_iU^\ast\}_{i\in I})$ are abelian von Neumann algebras whose only common projections are $0$ and $I$, so their intersection is $\Co I$ by Proposition \ref{stonean completion}. 
\end{proof}



As we have demonstrated now, that not only is the Hilbert Lattice we originally embedded $CL$ into minimal, the von Neumann algebra as a whole is generated by two copies of CL with orthogonal atoms. 

\begin{coro}
Let $L(C)$ be the meet semi lattice generated by the $C$ the coatoms of the cubic lattice adjoin $1$. Then the meets and joins of $L(S)$ are exactly $CL \subseteq B(H).$ 
\end{coro}

We can see that our generation of $B(H)$ is therefore a generalization of the single qubit case to arbitrary cardinals. 



Now we have shown that $B(H)$ is generated by $\De$, $CL$ directly. We also see that $B(H)$ is a minimal structure containing both, and as such is a necessary structure if one considers an operator algebraic structure of the cubic lattice under the conditions \cite{qlog2} detailed at the conclusion of section 4.1. 

\subsection{Phase Rotations}
So far we have re-derived the Pauli and Hadamard gates, referred to as the X, Z, and H gates in the literature, and their respective role in the underlying von Neumann algebra. As shown this von Neumann algebra is over a Hilbert space constructed in the standard manner and generalized to arbitrary cardinals. The question now becomes what types of observables can we obtain as functions of our already constructed observables? We will show that continuous functional calculus can be used to construct universal quantum gates in the sense of the Solovay Kitaev theorem, \cite{universal}. 

\begin{defin}
Let $U_\De$, $s$ be represented in $M_2(\Co)$, then 
\begin{align*}
R_x(\theta) &=  e^{i  U_\De \dfrac{\theta}{2}}  = 
\begin{bmatrix}
cos\left(\dfrac{\theta}{2}\right) & -i sin\left(\dfrac{\theta}{2}\right) \\
-isin\left(\dfrac{\theta}{2}\right) & cos\left(\dfrac{\theta}{2}\right) \\
\end{bmatrix}\\
R_y(\theta) &= e^{iU_\De s \dfrac{\theta}{2} } = 
\begin{bmatrix}
cos\left(\dfrac{\theta}{2}\right) & - sin \left(\dfrac{\theta}{2}\right) \\
sin\left(\dfrac{\theta}{2}\right) & cos\left(\dfrac{\theta}{2}\right) \\
\end{bmatrix}\\
R_z(\theta) &= e^{i s \dfrac{\theta}{2}} = 
\begin{bmatrix}
e^{\dfrac{-i \theta}{2}} & 0 \\
0 & e^{\dfrac{i\theta}{2}} \\
\end{bmatrix}\\
\end{align*}
\end{defin}

We will discuss group theoretic properties that can be shown directly from a computation in the case of $M_2(\Co)$, but we highlight a more general, standard technique to extend these results.

\begin{prop}\label{unitary similiarity and continuous functions}
Let $A$ be a normal operator in a $C^\ast$ algebra, $\mathscr{A}$. Then for any $f \in C(\sigma(A))$, and unitary $U \in \mathscr{A}$,  $Uf(A)U^\ast = f(UAU^\ast)$. 
\end{prop}

\begin{proof}
Let $\rho(1) = I$, $\rho(z) = A$ be the standard continuous functional calculus on $A$. Let $\gamma = U\rho U^\ast$, and let $\tau(1) = I$, $\tau(z) = UAU^\ast$. As a transformation by unitary similarity does not change the spectrum of $A$, our mappings, $\gamma$ and $\tau$, both have the same domain $C(\sigma(A))$. We have $\gamma(1) =  U\rho(1)U^\ast = UIU^\ast = I = \tau(1)$, and $\gamma(z) = U\rho(z)U^\ast = UAU^\ast = \tau(z)$, and the result follows for any continuous function by the uniqueness of the continuous functional calculus.
\end{proof}


\begin{lemma}
Let $U$ be a unitary operator and $A$ be a normal operator in a $C^\ast$ algebra such that $UA = -AU$. Then for any $t \in \Co,$ $Ue^{tA}U^\ast = e^{-tA}$.
\end{lemma}

\begin{proof}
We apply Proposition \ref{unitary similiarity and continuous functions} to see that $Ue^{tA}U^\ast = e^{tUAU^\ast} = e^{-tA}$.
\end{proof}

Now we can use the above lemmas to immediately deduce that the action of unitary similarity of any member of a Cartesian triple acts as inversion of rotation of any other member of the same Cartesian triple. Explicitly, $e^{x} e^{-x} = 1$ when considered as standard continuous functions over $\Co$, and we have an algebra homomorphism for the respective operator valued functions.  Furthermore, the action of unitary similarity of any normal element on its own exponent function is trivial.

\begin{theorem}
Let $G = \lan U_\De,$ $e^{2 \pi \theta i s_i} \ran$. Then $G \cong D_{2n}$ for some $n \in \N$, if $\theta$ is a rational  or $D_\infty$ if $\theta$ is an irrational.
\end{theorem}

\begin{proof}
We recognize from the above discussion that $U_\De$ embeds to the automorphism group generated by $e^{2 \pi \theta i s_i}$ as inversion, so we take the semidirect product. With the presentation $\lan U_\De, e^{2 \pi \theta i s_i} : U_\De e^{2 \pi \theta i s_i} U_\De = e^{-2 \pi \theta i s_i } \ran$,  and we see that the isomorphism type of the group follows by the order of $e^{2 \pi \theta i s_i}$, which is finite if $\theta$ is a rational and infinite otherwise, so the result follows. 
\end{proof}

\begin{coro}
Let $G = \lan U_\De, e^{i \sum_{i \in I} 2\pi \theta_i s_i} \ran$. Then $G \cong D_{2n}$ for some $n \in \N$ if $\theta = 1$ for  all but finitely $i \in I$ and $\theta_i \in \mathbb{Q}$ for finite $i$, or $G \cong D_\infty$ otherwise.
\end{coro}

\begin{proof}
We need only apply the previous theorem to each $s_i$ and use that continuous functions of commuting operators commute by functional calculus. If there are only finitely many rational $\theta$ not equal to one, then we can consider the lcm of their respective orders to obtain a finite $n$ satisfying the claim. 
\end{proof}

We now compare the above representation to the universal representation. 

\begin{coro}
Let $\mathscr{A} = C^\ast(D_{2n})$, $3 \le n \in \N$ in the representation $\pi: G \r B(H)$, where $H$ is of Theorem \ref{embedding}, and $\mathscr{B}$ be the reduced $C^\ast$ algebra of $D_{2n}$ with left regular representation $\lambda : G \r B(l^2(G))$. Then $\mathscr{A}$ is a nontrivial quotient of $\mathscr{B}$.
\end{coro}

\begin{proof}
We start with $n \ge 4$ and assume that $\theta_i = 1$ for all but exactly one $k \in I$. Without loss of generality, we assume $k = 1$.  As $D_{2n}$ is a group extension of discrete groups, $D_{2n}$ is amenable, and we have that the reduced $C^\ast$ algebra and the universal $C^\ast$ algebra are isomorphic, so we only need to show that $||\pi(a)|| < ||\lambda(a)||$ for some $a \in \mathscr{A}$. Let us consider the group ring $\Co[G]$ and restrict to elements over the cyclic subgroup $\Z_n \cong \lan r \ran$ in each representation. 

Then $\lambda(\sum_{j = 0}^{n-1} c_j r^j) \ne 0$ for any choice of $c_j \in \Co$ as the $r_j$ are linearly independent. However, $\pi(r) = R \otimes (\otimes_{i \in \{I  -1\}} I_2)$, for an appropriate rotation matrix $R \in M_2(\Co)$, and as a vector space, $C^\ast(R)$ has dimension at most 3 because $U_\De \notin C^\ast(R)$ as $U_\De$ does not commute with $R$ and $C^\ast(R)$ is an abelian algebra. Therefore, $\pi(\sum_{j = 0} ^{n-1} c_j r^j) = 0$ for some choice of $c_j \in \Co$.

Now let $n = 3$, we can directly compute that $\pi(a) = I + R + R^2 = 0$, so $\pi(a) = 0 < ||\lambda(a)||$, again using linear independence. 

We have shown the result for a single coordinate of the tensor product, and if we extend to the multi-coordinate tensor case, we have for some element $a \in \mathscr{A}$, $\pi(a_i) < \lambda(a_i)$, so the same must be true for the product of the norms across the indexing set. 
\end{proof}


\begin{remark}
We want to highlight that this behavior is quite different when considering the relation of anti-commutativity of the product of $s_i$. $U_\De$ and $\Pi_{j \in J}$ $s_i$ anti commute exactly when $J$ is odd, and the relationship is non-obvious when $J$ is infinite. This is because -1 factors through the tensor product, and we get a term $(-1)^n$ as a leading coefficient. However, as described above this does not occur when we consider exponentiation of the respective product. 
\end{remark}

We conclude having demonstrated that many of the \enquote{classical} quantum gates are a direct consequence of our construction of the cubic lattice as an orthomodular lattice of orthogonal projections. Due to our construction we have another natural choice of representation in a more geometric view as a cube in dimension $|I|$ as opposed to the larger $2^{|I|}$, which may have interesting application on its own. In addition, we have shown a number of group theoretic properties of their respective algebras, and that the remaining gates can be naturally constructed as functions of the already constructed gates both in a direct sense via the exponential map, and in a more general sense as an observable constructed of a continuous or Borel function over the spectrum of a Cartesian triple using the spectral mapping theorems. From a physical perspective, we have given a  mathematically formal description of the lattice of observables for a system of spin -$\frac{1}{2}$ of arbitrary cardinal. Furthermore, the gates $I$, $De$, $\sqrt{s}$, $U$, or more standardly $I$, $X$, $\sqrt{Z}$, $H$ combined with classical circuits generate a set of universal quantum gates. 

\printbibliography[
heading=bibintoc,
title={References}
]
\end{document}